\documentclass[twocolumn,showpacs,preprintnumbers,draftclsnofoot]{revtex4}
\usepackage{amsmath}
\usepackage{amssymb}
\usepackage{graphicx}
\usepackage{dcolumn}
\usepackage{bm}
\usepackage{amssymb}
\usepackage{graphicx}
\usepackage{dcolumn}
\usepackage{bm}

\begin{document}

\title{ Large photon drag effect of intrinsic graphene induced by plasmonic evanescent field}
\author{Ma Luo and Zhibing Li\footnote{Corresponding author:stslzb@mail.sysu.edu.cn}}
\affiliation{The State Key Laboratory of Optoelectronic Materials and Technologies \\
School of Physics\\
Sun Yat-Sen University, Guangzhou, 510275, P.R. China}

\begin{abstract}

Large photon drag effect of the massless Dirac Fermions in intrinsic graphene is predicted for a graphene-on-plasmonic-layer system. The surface plasmons in the plasmonic layer enlarge the wave number of photon for hundreds of time of that in vacuum. The evanescent field of the surface plasmons generates directional motion of carriers in the intrinsic graphene, because of the large momentum transfer from the surface plasmon to the excited carriers. A model Hamiltonian is developed on the assumption that the in-plane wavelength of the surface plasmons is much smaller than the mean free path of the carriers. The time evolution of density matrix is solved by perturbation method as well as numerical integration. The non-diagonal density matrix elements with momentum transfer lead to gauge current that is the optically driven macroscopic direct current. The dependence of the macroscopic direct current on the incident direction and intensity of the laser field is studied.
\end{abstract}

 \maketitle

\section{Introduction}

Graphene, a two dimensional mono-atomic carbon layer in honey-comb lattice \cite{neto09,basov14}, is a novel material for optoelectronic applications. \cite{Xuetao13,Gosciniak13,RanHao13,WeiLi14} Numerous researches on optical excitation and carrier kinetics of graphene have been reported. Investigations on linear responds of graphene to optical field reveal the dispersive conductivity of graphene. \cite{hsu07,hwang07,Falkovsky07,Stauber08,Wenhu08,Berahman15} Doped graphene with Fermi level away from the Dirac point is found to support the surface plasmons(SPs). \cite{jablan09,jablan13} Intrinsic graphene is found to have constant absorption coefficient for optical field with frequency up to infrared and intensity below the lower bound of the nonlinear optical region. Carrier kinetics under excitation of optical pulses with intensity in the linear \cite{Breusing11,malic11,Sun13,Torben14} and the nonlinear \cite{Mkrtchian12,Mkrtchian13,sipe11,sipe15,kelardeh15} optical regions have been investigated by semiconductor Bloch equations. The excited carrier thermal relaxations due to electron-phonon and electron-electron scatterings, are effective at the time scale of picosecond and femtosecond, respectively. \cite{malic11} Oblique incident continue wave laser induced direct current, which is called photon drag effect, has been investigated. \cite{Entin10,Karch10,Chongyun11,Glazov14} In addition, the linear photogalvanic effect is found in graphene on subtract that breaks the centrosymmetric.

The previous investigation of photon drag effect considered an oblique incident laser beam from vacuum. The direct current induced by momentum transfer directly from photon to the carrier, which is called non-resonant photon drag effect, is small due to small value of photon momentum in vacuum. For resonant photon drag effect, the resonant excited electrons above the Fermi level and the holes in the valence band form a net flux of charge that carries larger photon drag current. \cite{Entin10} Because the in-plane wave length of the incident field is larger than the mean free path in the case of \cite{Entin10,Karch10,Chongyun11,Glazov14}, the carriers were treated as quasi-classical particles that are driven by the electrical and Lorentz force. The Boltzmann equation was solved to obtain the second order conductivity that is in accordance with the parameter in phenomenological description of the photon drag effect. \cite{Karch10,Chongyun11}

In this paper, we consider the photon drag effect of intrinsic graphene in a graphene-on-plasmonic-layer structure. The plasmonic layer supports SPs with in-plane wave number hundreds of time larger than the wave number in vacuum, and confines the optical field in sub-wavelength region. \cite{jablan09,jablan13,rana11,yuan11} The local enhancement of optical field intensity near plasmonic layer enhances nonlinear optical response of graphene, such as second harmonic generation. \cite{Mikhailov11,Smirnova15} In our proposed structure, the photon drag effect is enhanced by the evanescent field of the SP mode that contains greatly enlarged photon momentum. In our case the in-plane wave length is much smaller than mean free path of carriers, therefore the nonlocal nature of electron wave function disqualifies the quasi-classical particle picture. In order to describe this photon drag effect, we use the quantum theory of optical excitation and generalize the semiconductor Bloch equations. The intrinsic graphene assumes symmetric behavior of electron and hole, so that the photocurrent under the consideration is due to non-resonant photon drag effect.

The article is organized as following: In section II, the physical model of the photon drag effect with large photon momentum transfer is explained, and a graphene-on-plasmonic-layer system is proposed to implement this effect. In section III, the theoretical and numerical study is presented. The phenomenological description explains the property of the photon drag effect under the excitation of p polarization evanescent field. The origin and properties of the photon drag current as well as the asymmetric excitation of carriers are discussed. In section IV, the conclusion is given.

\section{physical model}

\begin{figure}[tbp]
\scalebox{0.5}{\includegraphics{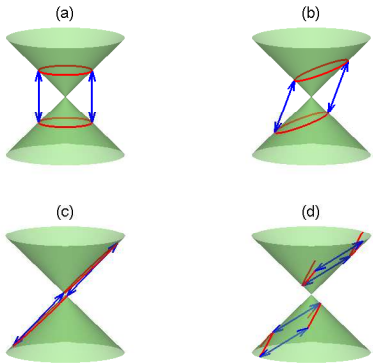}}
\caption{ The schematic picture
for the optical transitions in the Dirac cone of a 2D massless Dirac fermions system, obeying energy and momentum conservation. In (a) the in-plane momentum of
photon is zero. In (b), (c) and (d) the in-plane momentum of photon is non-zero, and phase velocity of the photon $\omega/|\mathbf{q}|$ is larger than, equal to, and smaller than that of the massless Dirac fermions $v_{F}$, respectively. }
\label{fig_cone}
\end{figure}

The low energy excitations of graphene having wave vectors near to two Dirac
points, $\mathbf{K}=(\frac{4\pi}{3\sqrt{3}a},0)$ and $\mathbf{K}^{\prime}=(-\frac{4\pi}{3\sqrt{3}a},0)$ in the $(k_{x},k_{y})$ plane, are described by massless Dirac fermions that have the linear dispersion, with $a=0.142nm$ being the bond length between two neighbored carbon atoms. In the vicinity of $\mathbf{K}$ for instance, $\varepsilon_{\mathbf{k}}^{\lambda}=\lambda\hbar
v_{F}|\mathbf{k}-\mathbf{K}|$, with $\lambda=\pm1$ for the conduction band and the valence
band respectively, $v_{F}\approx c_{0}/330$ being the Fermi velocity, and $c_{0}$
being the speed of light in vacuum. \cite{neto09,basov14} Compared to the dispersion of
photon in dielectric material with refraction index $n$ and wave vector $\mathbf{q}$,
$\hbar\omega=\hbar c|\mathbf{q}|/n$, it can be seen that for the same energy, the momentum of the
electron is $330/n$ times larger than that of photon. For the normal incident laser, the in-plane wave vector vanishes, and the electron momentum conserves. \cite{malic11,Torben14,Mkrtchian12,sipe11,kelardeh15,kumar16} This type of optical excitation is shown in Fig.
\ref{fig_cone}(a) where the lower(upper) cone is the valence(conduction) band, the vertical blue lines with double arrows imply transitions with vanishing momentum transfer. For the oblique incident laser, which is the case of photon drag effect, the excited electrons gain momentum. This type of excitation is shown in Fig. \ref{fig_cone}(b), where the tilted angle of the blue lines is determined by the ratio between the photon energy and momentum. Because $n$ is smaller than 10 for most dielectric media, the blue lines should be nearly vertical. We explore the plasmonic system that effectively produces $n\approx300$, so that the tilted angle of the blue lines in the excitation picture is large, as shown in Fig. \ref{fig_cone}(b-d).

\begin{figure}[tbp]
\scalebox{0.21}{\includegraphics{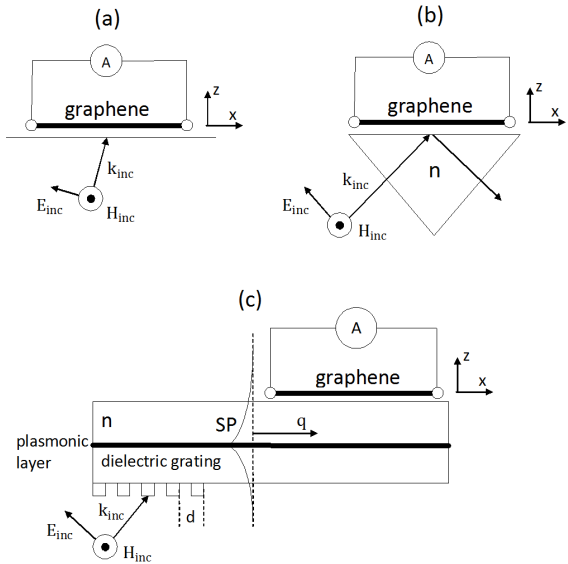}}
\caption{ Sketch of optical system exhibiting photon drag effect induced by traveling plane wave in (a), by evanescent field of total reflection at the interface in (b), and by the evanescent field of SPs of the graphene-on-plasmonic-layer system in (c). Dielectric grating in (c) converts the incident plane wave to the SP of the plasmonic layer. } \label{fig_system}
\end{figure}

A regular system exhibiting the photon drag effect is shown in Fig. \ref{fig_system}(a), where the traveling wave shines on the graphene and transfers momentum to electrons of graphene directly. Fig. \ref{fig_system}(b) is an alternative set up that the evanescent field produced by the total reflection of incident light excites electrons of graphene near to the reflection surface. The in-plane photon wave number is $2\pi n\sin(\theta_{inc})/\lambda_{0}$, with $\lambda_{0}$ being the wavelength in vacuum, and $\theta_{inc}$ being the incident angle. The graphene-on-plasmonic-layer system proposed in the present paper is given in Fig. \ref{fig_system}(c). The plasmonic layer being embedded in the vicinity of the dielectric surface could be consisting of doped graphene \cite{jablan13,jablan09,rana11}, mono-atomic layer of silver atoms \cite{Nagao01}, or aluminum atoms \cite{yuan11}. These types of plasmonic layers support SP modes with in-plane wave number being $2\pi n_{SPP}/\lambda_{0}$ and $n_{SP}\approx300$. Excitation of SP modes by the incidence of p polarization plane wave at a dielectric grating requires quasi-phase matching, $2\pi n\sin(\theta_{inc})/\lambda_{0}+2\pi N/d=2\pi n_{SPP}/\lambda_{0}$, with $d$ being the period of the grating and $N$ being an integer. The dielectric grating can be generated by optoacoustics grating with $d$ being determined by the acoustic wavelength. \cite{farhat13,schiefele13} Direct fabrication of the dielectric grating by etching the dielectric substrate is possible as well. \cite{Xiaolong13} The SPs propagate in the sub-wavelength region, with the squeezed E.M. field (evanescent field) confined in the vicinity of the surface, decreasing at the out-of-plane direction exponentially. \cite{christensen12,nikitin11,nikitin12} The plasmonic system strongly enhances the light-matter interaction because of highly localization of the optical field. \cite{koppens11} The intrinsic graphene deposits on the insulating dielectric surface with several nanometers in separation from the plasmonic layer. Note that the intrinsic graphene does not support large wave number non-over-damping SP mode by its self at room temperature \cite{Sarma13}, so that electrons in the intrinsic graphene are only excited by the evanescent field generated by of the plasmonic layer.

The evanescent field above the plasmonic layer is
\begin{equation}
\mathbf{E}=E_{0}e^{i\mathbf{q}\cdot\mathbf{r}-i\omega t}e^{-q_{z}z}(\frac{1}{\sqrt{2}}\hat{\mathbf{q}}+\frac{i|\mathbf{q}|}{\sqrt{2}q_{z}}\hat{z})+c.c. \label{electricField}
\end{equation}
where $E_{0}$ is the electric field amplitude on the plasmonic layer, $\mathbf{q}$ is the in-plane wave vector, $\mathbf{r}$ is the in-plane spatial coordinate, $\omega$ is the frequency of the SP mode, $q_{z}=\sqrt{|\mathbf{q}|^{2}-(\omega/c_{0})^2}$ is the decay rate, $z$ is the vertical distance from the plasmonic layer, $\hat{\mathbf{q}}$ and $\hat{z}$ are the in-plane and out-of-plane unit vectors, respectively. Because $|\mathbf{q}|\approx300\omega/c_{0}$, we have $|\mathbf{q}|/q_{z}\approx1$. Applying energy and momentum conservation, the allowed states excited by this E.M.
field are shown in Fig. \ref{fig_cone}(b) to (d) with red curves.  For the SP mode with phase velocity $\omega/|\mathbf{q}|$ larger than $v_{F}$, the excitation process is shown in
Fig. \ref{fig_cone}(b), which is an interband transition.
Comparing to the regular optical excitation of Fig. \ref{fig_cone}(a),
the distribution of excited electrons (the upper red circle) and that of excited holes (the lower red circle) are asymmetric in the 2D reciprocal space. More
forward electron states are excited than backward states, relative to
the propagation direction of the SP mode. When the phase velocity of the SP mode approaches
$v_{F}$, the allowed transitions are shown in Fig.
\ref{fig_cone}(c). In this case,
forward (backward) moving electrons (holes) are dominating. The phase volume of the states involved in the allowed transitions, which
is the length of the red line in Fig. \ref{fig_cone}(c), has the same order of magnitude as
that of the regular optical excitation of Fig. \ref{fig_cone}(a).
This is a specific property of massless Dirac fermion
systems. For 2D non-relativistic electron gas, in contrast, the dispersion is
parabolic therefore the phase volume of the excited states shrinks to
zero as the phase velocity of the SP mode decreases. Therefore the non-relativity electron system is not feasible for the SP excitation. For the SP mode with phase velocity smaller than $v_{F}$, only
intraband transition is possible, as shown in Fig.
\ref{fig_cone}(d) by the red curves. Because the valence band is fully filled for intrinsic
graphene, this type of transition is negligible.

\begin{figure}[tbp]
\scalebox{0.4}{\includegraphics{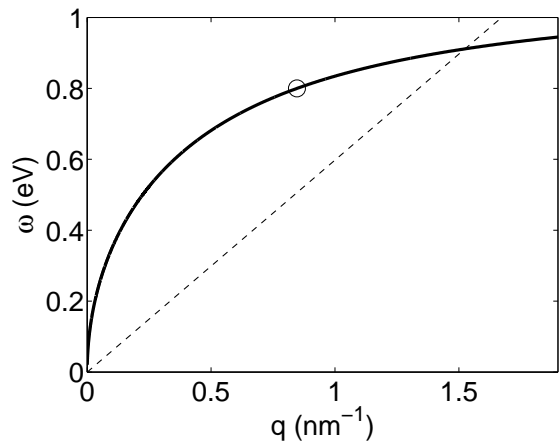}}
\caption{ The typical dispersion of the SP mode of plasmonic layer(solid line) and the dispersion of the massless Dirac fermions of the graphene(thin dashed line). The circular point marks the SP mode that we choose for the numerical calculation. } \label{fig_dispersive}
\end{figure}

In our specific model, the plasmonic layer that supports the SP mode consists of doped graphene with Fermi level being 0.66
eV. The dielectric substrate is $SiO_{2}$ with permittivity being 1.5. The plasmonic layer is imbedded 8nm below the top surface. The separation between the graphene sheet on the top and the plasmonic layer is large enough such that their electronic states are un-coupled. The  SP field is localized near to the plasmonic layer with decay length around 1.2 nm. Thus, in the calculation of the dispersion of the SP mode, the boundary effect of the substrate can be neglected. The dispersive curve of the SP is plotted in Fig. \ref{fig_dispersive} as solid line. \cite{jablan13,jablan09} We choose the SP mode with frequency being 0.8eV and wave number being $0.85 nm^{-1}$, which has a large momentum as well as a long propagation length. The dispersion of the massless Dirac Fermion of intrinsic graphene is also plotted as the thin dashed line for comparison. The phase velocity of this SP mode is larger than $v_{F}$, so that the optical transition is corresponding to the case of Fig. \ref{fig_cone}(b). In order to have the optical transitions shown in Fig. \ref{fig_cone}(c) and (d), other 2D materials that support SP modes with smaller phase velocity is needed, and that is out of the scope of this article.

\section{theoretical model and numerical result}

\subsection{Phenomenological description of photon drag effect}

The photon drag effect and photogalvanic effect are the second order nonlinear optical phenomena that are phenomenologically determined by the second order susceptibility. For ideal graphene with $D_{6h}$ point group, there are only four independent components in the second order susceptibility tensor. The electric field polarization vector of the SP is $\mathbf{e}=e_{q}\hat{\mathbf{q}}+e_{z}\hat{z}$ with $\hat{\mathbf{q}}$ and $\hat{z}$ being defined by Eq. (\ref{electricField}). The photon drag current is parallel to $\mathbf{q}$, and is given as
\begin{equation}
\mathbf{j}\cdot\hat{q}=(T_{1}+T_{2})|\mathbf{q}|\frac{|e_{q}|^{2}}{2}I+T_{4}|\mathbf{q}||e_{z}|^{2}I
\end{equation}
where $T_{i}$(i=1,2,3,4) are the four independent components of the second order susceptibility tensor, $I=|E_{0}|^{2}(c_{0}/2\pi)$ is the optical intensity. The term related to $T_{3}$ vanishes because the spatial derivatives of both $\mathbf{E}$ and $\mathbf{E}^{*}$ with respect to z coordinate produce the same factor $-q_{z}$ for the evanescent wave. The term related to $T_{4}$ is proportional to the thickness of the graphene, which makes this term negligible. Because the graphene deposits on the dielectric substrate that break the symmetry of up and down, the non-centrosymmetric background might induce the photogalvanic effect. The photogalvanic current is given as
\begin{equation}
\mathbf{j}\cdot\hat{q}=\chi_{l}\frac{e_{q}e_{z}^{*}+e_{q}^{*}e_{z}}{2}I
\end{equation}
where $\chi_{l}$ is the nonzero component of the second order susceptibility. Because $e_{z}/e_{q}$ is imaginary for the evanescent field, the photogalvanic current vanishes. In summary, the direct current generated by the evanescent field is solely the photon drag effect, with the current direction parallel to $\mathbf{q}$ and the amplitude being $(T_{1}+T_{2})|\mathbf{q}|I/4$.

\subsection{Model Hamiltonian}

Previous investigation of the photon drag effect has treated the spatial inhomogeneous part of the interaction Hamiltonian by the first order spatial correction. The spatial exponential factor of the electric field is expanded as $e^{i\mathbf{q}\cdot\mathbf{r}}\approx1+i\mathbf{q}\cdot\mathbf{r}$. This treatment is valid for the case that the in-plane wavelength $2\pi/|\mathbf{q}|$(or $\lambda_{0}/\sin\theta_{inc}$ for the oblique incident plane wave) is larger than the mean free path of the carriers in graphene. Correspondingly, the carriers have been treated as classical particles in the previous study. The distribution of electron in the real and reciprocal space is described by the Boltzmann equation. When the in-plane wavelength $2\pi/|\mathbf{q}|$ is much smaller than the mean free path, the spatial exponential factor of the interaction Hamiltonian should be kept. In our specific example, the in-plane wavelength of the SP mode is equal to 7.4 nm that is much smaller than the mean free path of about 1 $\mu m$ for graphene\cite{neto09}. Instead of Boltzmann equation, we use the semiconductor Bloch equations to describe the distribution of carriers in the reciprocal space.

The electron in graphene is modeled by the tight binding theory,  which
gives the wave functions of the non-interacting eigenstates
$\left|\lambda\mathbf{k}\right\rangle$ in the real space as
\begin{equation}
\Psi_{\lambda}(\mathbf{k},\mathbf{r})=\sum_{s=A,B}{C_{\lambda}^{s}(\mathbf{k})\frac{1}{\sqrt{N}}\sum_{\mathbf{R}_{s}}{e^{i\mathbf{k}\cdot\mathbf{R}_{s}}\phi(\mathbf{r}-\mathbf{R}_{s})}}
\label{noninter eigenstate}
\end{equation}
with $\mathbf{R}_{s}$ being the lattice vectors of the A and B
atoms, and $\phi(\mathbf{r}-\mathbf{R}_{s})$ being the spatial wave function of the $2p_{z}$ orbital at the lattice site $\mathbf{R}_{s}$. The compound index of the eigenstates contains $\lambda=+1$($-1$) standing for conduction(valence) band, and Bloch wave vector $\mathbf{k}$. The energy levels and coefficients $C_{\lambda}^{s}$ are obtained by diagonalizing the Hamiltonian under the tight binding basis \cite{neto09,basov14}.

The interaction Hamiltonian for electron in the evanescent field of the SP mode
is,
\begin{equation}
H_{I}=-\frac{e_{0}}{m_{0}c}\mathbf{A}\cdot\mathbf{P}=i\frac{\hbar
e_{0}}{m_{0}c}\mathbf{A}\cdot\nabla
\end{equation}
where $\mathbf{A}$ is the vector potential. Under the Coulomb gauge, $\mathbf{E}=-\frac{1}{c_{0}}\frac{\partial\mathbf{A}}{\partial t}$. $\mathbf{E}$ is the electric field of Eq. (\ref{electricField}) multiplied by a slow varying profile function $f(t)$. The interaction between electron and the $\hat{z}$ component of the electric field of the SP mode is
neglected because of its small effect on the optical excitation of
graphene. For a
general SP wave package, such as the Gaussian pulse, multiple modes with various $\mathbf{q}$
should be included. In order to avoid the computational complexity, only SP of single mode will be considered. In the present work, the time profile function $f(t)$ is chosen to be the hyperbolic tangent function starting from zero, and with a turn-on time much larger than the period of the SP mode. Using the slow varying approximation that assumes $f'(t)<<\omega$ and $Im[f(t)]=0$,
the interaction Hamiltonian is given as,
\begin{eqnarray}
H_{I}&=&-i\frac{\hbar
e_{0}}{m_{0}}\int_{0}^{t}{\mathbf{E}(t')dt'}\cdot\nabla \nonumber \\
&\approx&\frac{\hbar
e_{0}E_{0}}{2m_{0}\omega}f(t)(e^{i\mathbf{q}\cdot\mathbf{r}-i\omega
t}-e^{-i\mathbf{q}\cdot\mathbf{r}+i\omega t})\hat{\mathbf{q}}\cdot\nabla
\end{eqnarray}
Because of momentum conservation, the transition matrix elements are nonzero only if the initial and final Bloch wave vectors are different by $\pm\mathbf{q}$. Thus, the non-zero matrix elements
of the SP absorption process are written as,
\begin{eqnarray}
&&\left\langle\lambda'\mathbf{k}+\mathbf{q}\right|H_{I}\left|\lambda\mathbf{k}\right\rangle= \nonumber \\
&~~~~~&\frac{\hbar
e_{0}E_{0}f(t)}{2m_{0}\omega}  \left\langle\lambda'\mathbf{k}+\mathbf{q}\right|e^{ i\mathbf{q}\cdot\mathbf{r}}\hat{\mathbf{q}}\cdot\nabla\left|\lambda\mathbf{k}\right\rangle e^{-i\omega
t} \label{matrixab1}
\end{eqnarray}
with
\begin{eqnarray}
& &\left\langle\lambda'\mathbf{k}+\mathbf{q}\right|e^{ i\mathbf{q}\cdot\mathbf{r}}\hat{\mathbf{q}}\cdot\nabla\left|\lambda\mathbf{k}\right\rangle
= \nonumber \\ &~&m\sum_{i=1}^{3}{\frac{\mathbf{b}_{i}\cdot\mathbf{q}}{|\mathbf{b}_{i}|}
[C_{\lambda'}^{A*}(\mathbf{k}+\mathbf{q})C_{\lambda}^{B}(\mathbf{k})e^{i\mathbf{k}\cdot\mathbf{b}_{i}+\frac{1}{2}i\mathbf{q}\cdot\mathbf{b}_{i}}]}  \nonumber \\
&&- m\sum_{i=1}^{3}{\frac{\mathbf{b}_{i}\cdot\mathbf{q}}{|\mathbf{b}_{i}|}
[C_{\lambda'}^{B*}(\mathbf{k}+\mathbf{q})C_{\lambda}^{A}(\mathbf{k})e^{-i\mathbf{k}\cdot\mathbf{b}_{i}-\frac{1}{2}i\mathbf{q}\cdot\mathbf{b}_{i}}]}
\label{matrixab2}
\end{eqnarray}
where $\mathbf{b}_{i}$  ($i=1,2,3$) are three vectors from an A atom to its three nearest neighbored B atoms, $m\approx3nm^{-1}$ is the norm of the matrix element of the Laplace operator between two nearest neighbored $2p_{z}$ orbitals. \cite{malic11,hsu07,ashish03} The matrix elements for the corresponding emission process are given by the complex conjugation of (\ref{matrixab1}) and (\ref{matrixab2}).

\subsection{Evolution of the Density Matrix}

The interaction Hamiltonian couples eigenstates $\left|\lambda\mathbf{k}\right\rangle$ with eigenstates $\left|\lambda'\mathbf{k}+\mathbf{q}\right\rangle$ for the SP absorption process, and with eigenstates $\left|\lambda'\mathbf{k}-\mathbf{q}\right\rangle$ for the SP emission process.
Define
$\left\langle\lambda'\mathbf{k}\right|H_{I}\left|\lambda\mathbf{k}+\mathbf{q}\right\rangle=H_{\lambda'\mathbf{k},\lambda\mathbf{k}+\mathbf{q}}f(t)e^{i\omega
t}$ and
$\left\langle\lambda'\mathbf{k}+\mathbf{q}\right|H_{I}\left|\lambda\mathbf{k}\right\rangle=H_{\lambda'\mathbf{k}+\mathbf{q},\lambda\mathbf{k}}f(t)e^{-i\omega
t}$. Applying the Heisenberg equation of motion,
$i\hbar\partial_{t}\rho(t)=[\rho(t),H]$, under the basis of the non-interacting tight binding eigenstates, we obtain the generalized semiconductor Bloch equations. The time evolution equations of the diagonal terms of the density matrix are given as
\begin{eqnarray}
\hbar\frac{\partial}{\partial t}\rho_{\lambda\mathbf{k},\lambda\mathbf{k}}&=&-2Im[H_{\lambda\mathbf{k}+\mathbf{q},\lambda\mathbf{k}}\rho_{\lambda\mathbf{k},\lambda\mathbf{k}+\mathbf{q}}f(t)e^{-i\omega t}]  \nonumber \\
&
&-2Im[H_{\bar{\lambda}\mathbf{k}+\mathbf{q},\lambda\mathbf{k}}\rho_{\lambda\mathbf{k},\bar{\lambda}\mathbf{k}+\mathbf{q}}f(t)e^{-i\omega
t}] \nonumber \\
& &+2Im[H_{\lambda\mathbf{k},\lambda\mathbf{k}-\mathbf{q}}\rho_{\lambda\mathbf{k}-\mathbf{q},\lambda\mathbf{k}}f(t)e^{-i\omega t}]  \nonumber \\
& &+2Im[H_{\lambda\mathbf{k},\bar{\lambda}\mathbf{k}-\mathbf{q}}\rho_{\bar{\lambda}\mathbf{k}-\mathbf{q},\lambda\mathbf{k}}f(t)e^{-i\omega
t}] \label{BlochEq1}
\end{eqnarray}
where $\bar{\lambda}=-\lambda$ is the opposite band index. The time evolution equations of the non-diagonal density matrix elements between eigenstates with the same Bloch wave vector but of different bands are given as
\begin{eqnarray}
\hbar\frac{\partial}{\partial
t}\rho_{v\mathbf{k},c\mathbf{k}}&=&i(\varepsilon_{c\mathbf{k}}-\varepsilon_{v\mathbf{k}})\rho_{v\mathbf{k},c\mathbf{k}} \nonumber \\
&-&iH_{v\mathbf{k},v\mathbf{k}+\mathbf{q}}\rho_{c\mathbf{k},v\mathbf{k}+\mathbf{q}}^{*}f(t)e^{i\omega
t} \nonumber \\
&+&iH_{v\mathbf{k}+\mathbf{q},c\mathbf{k}}\rho_{v\mathbf{k},v\mathbf{k}+\mathbf{q}}f(t)e^{-i\omega
t} \nonumber \\
&-&iH_{v\mathbf{k},c\mathbf{k}+\mathbf{q}}\rho_{c\mathbf{k},c\mathbf{k}+\mathbf{q}}^{*}f(t)e^{i\omega
t} \nonumber \\
&+&iH_{c\mathbf{k}+\mathbf{q},c\mathbf{k}}\rho_{v\mathbf{k},c\mathbf{k}+\mathbf{q}}f(t)e^{-i\omega
t}
\nonumber \\
&+&iH_{v\mathbf{k}-\mathbf{q},c\mathbf{k}}\rho_{v\mathbf{k}-\mathbf{q},v\mathbf{k}}^{*}f(t)e^{i\omega
t} \nonumber \\
&-&iH_{v\mathbf{k},v\mathbf{k}-\mathbf{q}}\rho_{v\mathbf{k}-\mathbf{q},c\mathbf{k}}f(t)e^{-i\omega
t} \nonumber \\
&+&iH_{c\mathbf{k}-\mathbf{q},c\mathbf{k}}\rho_{c\mathbf{k}-\mathbf{q},v\mathbf{k}}^{*}f(t)e^{i\omega
t} \nonumber \\
&-&iH_{v\mathbf{k},c\mathbf{k}-\mathbf{q}}\rho_{c\mathbf{k}-\mathbf{q},c\mathbf{k}}f(t)e^{-i\omega
t} \label{BlochEq2}
\end{eqnarray}
The time evolution equations of the non-diagonal density matrix elements between two eigenstates with Bloch wave vectors being different by $\mathbf{q}$ are given as
\begin{eqnarray}
& &\hbar\frac{\partial}{\partial t}\rho_{\lambda\mathbf{k},\sigma\lambda\mathbf{k}+\mathbf{q}} \nonumber \\
&=&i(\varepsilon_{\sigma\lambda\mathbf{k}+\mathbf{q}}-\varepsilon_{\lambda\mathbf{k}})\rho_{\lambda\mathbf{k},\sigma\lambda\mathbf{k}+\mathbf{q}} \nonumber \\
&+&iH_{\lambda\mathbf{k},\sigma\lambda\mathbf{k}+\mathbf{q}}(\rho_{\lambda\mathbf{k},\lambda\mathbf{k}}-\rho_{\sigma\lambda\mathbf{k}+\mathbf{q},\lambda\mathbf{k}+\mathbf{q}})f(t)e^{i\omega
t} \nonumber \\
&+&iH_{\bar{\lambda}\mathbf{k},\lambda\mathbf{k}+\mathbf{q}}\rho_{\lambda\mathbf{k},\bar{\lambda}\mathbf{k}}f(t)e^{i\omega
t} \nonumber \\
&-&iH_{\lambda\mathbf{k},\sigma\bar{\lambda}\mathbf{k}+\mathbf{q}}\rho_{\sigma\bar{\lambda}\mathbf{k}+\mathbf{q},\sigma\lambda\mathbf{k}+\mathbf{q}}f(t)e^{i\omega
t} \label{BlochEq3}
\end{eqnarray}
with $\sigma=+$ and $\sigma=-$ for intraband and interband non-diagonal density matrix elements, respectively.
Equation (\ref{BlochEq1}-\ref{BlochEq3}) are the generalized optical Bloch equations.

Beside the optical excitation, Coulomb scattering and phonon scattering will redistribute the excited electrons and holes. \cite{malic11} We apply the relaxation time approximation for the scattering processes, so that the Bloch equations of each density matrix element has an additional decay term, $-\Gamma_{\lambda\mathbf{k},\lambda'\mathbf{k}'}(\rho_{\lambda\mathbf{k},\lambda'\mathbf{k}'}-\rho_{\lambda\mathbf{k},\lambda'\mathbf{k}'}^{0})$, whence the scattering processes are accounted. The scattering rate of the diagonal density matrix element is given by $\Gamma_{\lambda\mathbf{k},\lambda\mathbf{k}}=\Gamma_{e-e}+\Gamma_{e-ph}(|\varepsilon_{\lambda\mathbf{k}}|)$, where $\hbar/\Gamma_{e-e}=30fs$ is the scattering rate of Coulomb scattering between electrons, $\Gamma_{e-ph}$ is the scattering rate of electron-phonon scattering. The most effective electron-phonon scattering is due to the optical phonon at $\Gamma$ point, with phonon energy equal to $0.2eV$ and scattering rate being $1/1200ps^{-1}$. This scattering event is not effective unless the energy level of electron or hole deviates from the Fermi level for more than $0.2eV$. Thus, $\Gamma_{e-ph}$ equates to 0($1/1200ps^{-1}$) for $|\varepsilon_{\lambda\mathbf{k}}|<0.2eV$($|\varepsilon_{\lambda\mathbf{k}}|>0.2eV$). Because the Fermi level of the intrinsic graphene is zero, the scattering rate of electron and hole is symmetric. The scattering rate of non-diagonal density matrix element is given as $\Gamma_{\lambda\mathbf{k},\lambda'\mathbf{k}'}=(1/2)(\Gamma_{\lambda\mathbf{k},\lambda\mathbf{k}}+\Gamma_{\lambda'\mathbf{k}',\lambda'\mathbf{k}'})+\Gamma_{off}$, with $\hbar/\Gamma_{off}\approx500fs$ being the off diagonal dephasing rate. The initial state is in equilibrium whose density matrix has vanishing off-diagonal elements and diagonal elements following the Fermi-Dirac distribution at the room temperature. The temperature is assumed unchanged in the following time.

\subsection{Generation of Direct Current}

The line current density of the graphene sheet can be calculated by the expectation value of
the momentum operator, $tr(\rho\mathbf{j})$, which is given as
\begin{equation}
\mathbf{j}(t)=\frac{e_{0}}{2Sm_{0}}\sum_{\lambda,\lambda',\mathbf{k},\mathbf{k}'}
\left\langle\lambda\mathbf{k}\right|\mathbf{p}-e_{0}\mathbf{A}(t)\left|\lambda'\mathbf{k}'\right\rangle\rho_{\lambda'\mathbf{k}',\lambda\mathbf{k}}+c.c. \label{define_current}
\end{equation}
where $S$ is the graphene area. Inserting the noninteracting basis functions of the tight binding theory (\ref{noninter eigenstate}), we obtain the expression for the total current as
\begin{eqnarray}
&&\mathbf{j}(t)=\frac{2\hbar e_{0}}{Sm_{0}}\{2\sum_{\mathbf{k}}{Im[\mathbf{M}^{vc}(\mathbf{k})\rho_{c\mathbf{k},v\mathbf{k}}]}  \nonumber \\
&&-i\sum_{\mathbf{k}}{[\mathbf{M}^{vv}(\mathbf{k})\rho_{v\mathbf{k},v\mathbf{k}}+\mathbf{M}^{cc}(\mathbf{k})\rho_{c\mathbf{k},c\mathbf{k}}]}\} \nonumber \\
&&-\frac{2e_{0}^{2}E_{0}c}{Sm_{0}\omega}\hat{\mathbf{q}}Im[\sum_{\lambda,\lambda',\mathbf{k}}{F_{\lambda'\mathbf{k}+\mathbf{q},\lambda\mathbf{k}}\rho_{\lambda\mathbf{k},\lambda'\mathbf{k}+\mathbf{q}}f(t)e^{-i\omega t}} ]  \nonumber \\
\end{eqnarray}
where
$\mathbf{M}^{\lambda\lambda'}(\mathbf{k})=\left\langle\lambda\mathbf{k}\right|\nabla\left|\lambda'\mathbf{k}\right\rangle$ is the in-plane-momentum conserving optical transition matrix and $F_{\lambda'\mathbf{k}+\mathbf{q},\lambda\mathbf{k}}=\left\langle\lambda'\mathbf{k}+\mathbf{q}\right|e^{ i\mathbf{q}\cdot\mathbf{r}}\left|\lambda\mathbf{k}\right\rangle$. The components of the current associated with $\mathbf{M}^{\lambda\lambda'}$ is the  microscopic canonical current, including interband current ($\mathbf{M}^{vc}$), and intraband currents ($\mathbf{M}^{vv}$ and $\mathbf{M}^{cc}$). The current component associated with $F_{\lambda'\mathbf{k}+\mathbf{q},\lambda\mathbf{k}}$ is the gauge current.

\subsection{Perturbation solution}
The steady state solution for $f(t)=1$ can be obtained by perturbation analysis. The zero order solution of the density matrix without the interaction Hamiltonian is the Fermi-Dirac distribution, denoted by $\rho^{(0)}$. The higher order solution is obtained by expanding the density matrix as $\rho=\rho^{(0)}+\delta\rho^{(1)}+\delta^{2}\rho^{(2)}...$. Inserting the expansions into the Heisenberg equation with interaction Hamiltonian(of order of $\delta$) and matching the coefficients with the same order of $\delta$, the first and second order perturbation solutions are obtained. The first order perturbation appears only in the non-diagonal density matrix elements between eigenstates with wave vectors different by $\mathbf{q}$, which are given as
\begin{equation}
\rho^{(1)}_{\lambda\mathbf{k},\lambda'\mathbf{k}+\mathbf{q}}=\frac{H_{\lambda\mathbf{k},\lambda'\mathbf{k}+\mathbf{q}}(\rho^{(0)}_{\lambda\mathbf{k}}-\rho^{(0)}_{\lambda'\mathbf{k}+\mathbf{q}})}{\hbar\omega+\hbar\omega_{\lambda\mathbf{k},\lambda'\mathbf{k}+\mathbf{q}}-i\Gamma_{\lambda\mathbf{k},\lambda'\mathbf{k}+\mathbf{q}}}e^{i\omega t}
\end{equation}
where $\hbar\omega_{\lambda\mathbf{k},\lambda'\mathbf{k}+\mathbf{q}}=\varepsilon_{\lambda\mathbf{k}}-\varepsilon_{\lambda'\mathbf{k}+\mathbf{q}}$. These terms induce the gauge current by coupling with the non-zero matrix element of the gauge field $F_{\lambda'\mathbf{k}+\mathbf{q},\lambda\mathbf{k}}$. The time oscillating factor is cancelled, so that the gauge current is direct current. The second order perturbation changes the diagonal density matrix elements, given as
\begin{eqnarray}
&&\rho^{(2)}_{\lambda\mathbf{k},\lambda\mathbf{k}}=\frac{2}{\Gamma_{\lambda\mathbf{k},\mathbf{k}}} \\
&&[Im(H_{\lambda\mathbf{k},\lambda\mathbf{k}+\mathbf{q}}\tilde{\rho}^{(1)}_{\lambda\mathbf{k}+\mathbf{q},\lambda\mathbf{k}})+Im(H_{\lambda\mathbf{k},\bar{\lambda}\mathbf{k}+\mathbf{q}}\tilde{\rho}^{(1)}_{\bar{\lambda}\mathbf{k}+\mathbf{q},\lambda\mathbf{k}})\nonumber \\
&&-Im(H_{\lambda\mathbf{k}-\mathbf{q},\lambda\mathbf{k}}\tilde{\rho}^{(1)}_{\lambda\mathbf{k},\lambda\mathbf{k}-\mathbf{q}})-Im(H_{\bar{\lambda}\mathbf{k}-\mathbf{q},\lambda\mathbf{k}}\tilde{\rho}^{(1)}_{\lambda\mathbf{k},\bar{\lambda}\mathbf{k}-\mathbf{q}})]\nonumber
\end{eqnarray}
and non-diagonal density matrix elements between eigenstates with the same wave vector, given as
\begin{eqnarray}
&&\rho^{(2)}_{\lambda\mathbf{k},\bar{\lambda}\mathbf{k}}=\frac{1}{\omega_{\lambda\mathbf{k},\bar{\lambda}\mathbf{k}}-i\Gamma_{\lambda\mathbf{k},\bar{\lambda}\mathbf{k}}} \\
&&(H_{\bar{\lambda}\mathbf{k}+\mathbf{q},\bar{\lambda}\mathbf{k}}\tilde{\rho}^{(1)}_{\lambda\mathbf{k},\bar{\lambda}\mathbf{k}+\mathbf{q}}-H_{\lambda\mathbf{k},\bar{\lambda}\mathbf{k}+\mathbf{q}}\tilde{\rho}^{(1)}_{\bar{\lambda}\mathbf{k}+\mathbf{q},\bar{\lambda}\mathbf{k}}\nonumber \\
&& H_{\lambda\mathbf{k}+\mathbf{q},\bar{\lambda}\mathbf{k}}\tilde{\rho}^{(1)}_{\lambda\mathbf{k},\lambda\mathbf{k}+\mathbf{q}}-H_{\lambda\mathbf{k},\lambda\mathbf{k}+\mathbf{q}}\tilde{\rho}^{(1)}_{\lambda\mathbf{k}+\mathbf{q},\bar{\lambda}\mathbf{k}}\nonumber
\\
&&H_{\bar{\lambda}\mathbf{k}-\mathbf{q},\bar{\lambda}\mathbf{k}}\tilde{\rho}^{(1)}_{\lambda\mathbf{k},\bar{\lambda}\mathbf{k}-\mathbf{q}}-H_{\lambda\mathbf{k},\bar{\lambda}\mathbf{k}-\mathbf{q}}\tilde{\rho}^{(1)}_{\bar{\lambda}\mathbf{k}-\mathbf{q},\bar{\lambda}\mathbf{k}}\nonumber \\
&& H_{\lambda\mathbf{k}-\mathbf{q},\bar{\lambda}\mathbf{k}}\tilde{\rho}^{(1)}_{\lambda\mathbf{k},\lambda\mathbf{k}-\mathbf{q}}-H_{\lambda\mathbf{k},\lambda\mathbf{k}-\mathbf{q}}\tilde{\rho}^{(1)}_{\lambda\mathbf{k}-\mathbf{q},\bar{\lambda}\mathbf{k}})\nonumber
\end{eqnarray}
where $\tilde{\rho}^{(1)}$ is the first order perturbation without the time oscillating factor. Both $\rho^{(2)}_{\lambda\mathbf{k},\lambda\mathbf{k}}$ and $\rho^{(2)}_{\lambda\mathbf{k},\bar{\lambda}\mathbf{k}}$ are time independent and induce the microscopic canonical  current of each Dirac cone. The microscopic canonical  current of each Dirac cone contains components that are both parallel and perpendicular to $\mathbf{q}$. Because of the combination of the spatial inversion symmetry and electron-hole symmetry for the relaxation time, the microscopic canonical  current at $\mathbf{K}+\mathbf{k}_{D}$ and $\mathbf{K}'-\mathbf{k}_{D}$ has opposite direction, with $\mathbf{k}_{D}$ being the displacement of wave vector from Dirac point. Thus, the total canonical  current vanishes. Applying the same scheme to 2D system with a single Dirac cone such as surface state of topological insulator\cite{qi11}, one could obtain nonzero canonical  current. The second harmonic oscillating terms appear in the second order perturbation of the density matrix elements $\rho^{(2)}_{\lambda\mathbf{k},\lambda'\mathbf{k}\pm2\mathbf{q}}$, which make no contribution to the macroscopic current density. The current density defined in Eq. (\ref{define_current}) is the average of current over the area that is at the length scale of the mean free path and much larger than $1/|\mathbf{q}|$. The second harmonic part of the current density spatially oscillates with wave number being $2|\mathbf{q}|$. Thus, the average of the second harmonic part of the current over the macroscopic volume vanishes. Therefore, the total macroscopic current density only contains the direct current component from the gauge current, i.e., the photon drag current. The direction of the current is parallel to $\mathbf{q}$, which agrees with the phenomenological argument.

\begin{figure}[tbp]
\scalebox{0.5}{\includegraphics{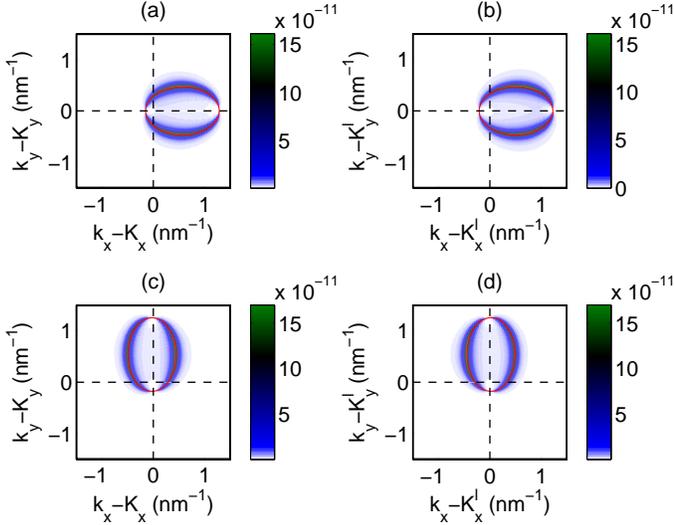}}
\caption{ (a) and (b) are population of excited electrons in conduction band near $K$ and $K'$ points, respectively, with the SP mode propagating along x axis or $\hat{q}=\hat{x}$. (c) and (d) are similar plots, with the SP mode propagating along y axis or $\hat{q}=\hat{y}$. The red curves indicate the states permitted by the theoretical selection rule of the optical excitation by the SP mode. } \label{fig_population}
\end{figure}

\begin{figure}[tbp]
\scalebox{0.5}{\includegraphics{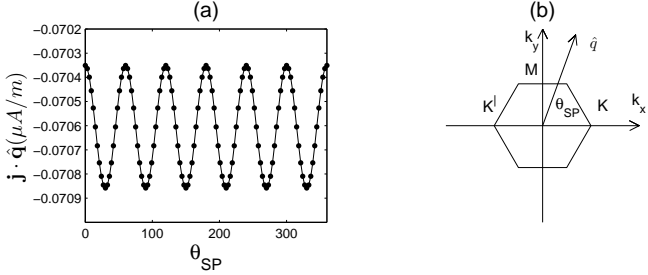}}
\caption{ (a) The optically excited current along the propagation direction of the SP mode(or $\hat{q}$), versus the propagation angle of the SP mode(the angle from $\hat{x}$ to $\hat{q}$). (b) indicates the direction of $\hat{q}$, and the $K$, $K^{'}$ and $M$ points in the reciprocal space.  }
\label{fig_DCcurrent}
\end{figure}

Because $H_{\lambda\mathbf{k},\lambda'\mathbf{k}+\mathbf{q}}$ and $\mathbf{A}$ are proportional to $E_{0}$, the magnitude of the photon drag current is proportional to $|E_{0}|^{2}$, for the second order perturbation solution. The photon drag effect induced by  a strong continue wave laser is calculated as example. The electric field amplitude at the center of the Gaussian beam of 9W, with beam width being $50\mu m$, is $0.93V/\mu m$. For simplicity we assume the dielectric grating has ideal transfer efficiency, thus the beam induces the SP mode with the same electric field $E_{0}$ at the plasmonic layer. Considering the out-of-plane decay factor of the evanescent field, the electric field at the graphene plane is $10^{-3}V/\mu m$. The population density of excited electrons in reciprocal space, i.e. $\rho_{c\mathbf{k},c\mathbf{k}}-\rho_{c\mathbf{k},c\mathbf{k}}^{0}$, are plotted in Fig. \ref{fig_population}. The directional excitation is exhibited in the excited electrons in two Dirac cones, which is consisting with the theoretical selection rules (red curves in Fig. \ref{fig_cone}) of energy and momentum conservation. However, not all of the selected states are excited equally because the transition amplitudes depend on the states and the propagation direction of the SP mode. The population of holes has similar distribution pattern as that of electrons, with a translation of $-\mathbf{q}$ in the reciprocal space. The amplitude of the photon drag current depends on the propagation angle of the SP mode(or $\theta_{SP}$), as shown in Fig. \ref{fig_DCcurrent}(a). $\theta_{SP}$ is defined as the angle from $\hat{x}$ to $\hat{q}$ as shown in Fig. \ref{fig_DCcurrent}(b). The direct current is symmetric under the rotation of $\hat{q}$ by $60^{o}$ because the hexagonal lattice of the graphene has the six-fold rotational symmetry.

\subsection{Numerical integration of the Bloch equations}

\begin{figure}[tbp]
\scalebox{0.4}{\includegraphics{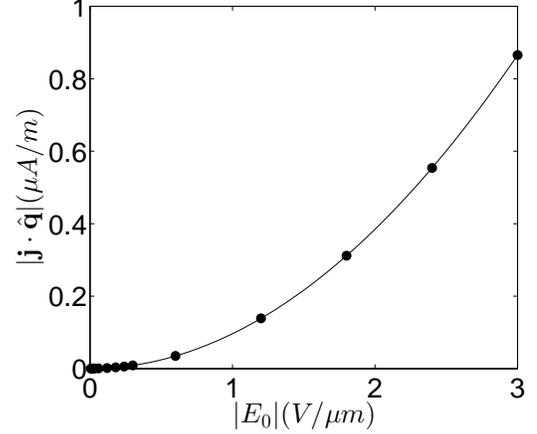}}
\caption{ The magnitude of the photon drag current versus the electric field magnitude at the plasmonic layer(dot). The solid curve is polynomial fit.  }
\label{fig_currentE0}
\end{figure}

The generalized Bloch equations can also be solved by numerical integration. We choose a slowly turn on function $f(t)$, and integrate the Bloch equations until that the diagonal elements become stable and the off-diagonal elements exhibit periodic behavior. The numerical result agrees well with the perturbation solution. The dependent of the current magnitude to the electric field, $|E_{0}|$, is calculated and plotted in Fig. \ref{fig_currentE0}. The polynomial fit of the numerical data shows that the photon drag current is proportional to $|E_{0}|^{2}$, or the local energy density of the evanescent field.

\section{conclusion}

In conclusion, we have proposed a graphene-on-plasmonic-layer system that can harness large momentum transfer through the SP mode and have large photon drag effect in the intrinsic graphene. The plasmonic layer supports the SP mode with large wave number along the in-plane propagation direction. Electrons in the intrinsic graphene parallel to the plasmonic layer are excited by the evanescent field of the SP mode. The excited electrons gain momentum along the propagation direction of the SP mode. Thus, the excited electrons and holes have asymmetric distributions in the reciprocal space. Base on the assumption that the wave length of the evanescent field is much smaller than the mean free path of the graphene, we developed the generalized semiconductor Bloch equations to describe the carrier dynamics in the graphene. The model is solved by perturbation method as well as numerical integration. The perturbation solution reveals that the photon drag current is originated from the gauge current. The microscopic canonical  currents of two Dirac cones are canceled by each other, because of the combination of the spatial inversion symmetry and electron-hole symmetry  for the relaxation time. The dependence of the magnitude of the photon drag current on the angle between the propagation direction of the SP mode and the $\mathbf{K}$ vector of graphene is calculated. The result manifests  the lattice symmetry of the graphene. The numerical solution of the Bloch equations confirms that the photon drag current is proportional to the square of the  electric field magnitude of the evanescent wave.

\begin{acknowledgments}
We thank H.J. Kreuzer for valuable comments. The project is supported by the National Natural Science Foundation of China (Grant:
11274393), the National Basic Research
Program of China (Grant: 2013CB933601), and the National Key Research and Development Project of China
(Grant: 2016YFA0202001).
\end{acknowledgments}

\clearpage


\section*{References}

\begin{thebibliography}{99}

\bibitem{neto09} A. H. Castro Neto, F. Guinea, N. M. R. Peres, K. S. Novoselov, and A. K. Geim, Rev. Mod. Phys., 81, 109(2009).

\bibitem{basov14} D. N. Basov, M. M. Fogler, A. Lanzara, Feng Wang, and Yuanbo Zhang, Rev. Mod. Phys. 86 959 (2014).

\bibitem{Xuetao13} Xuetao Gan, Ren-Jye Shiue, Yuanda Gao, Inanc Meric, Tony F. Heinz, Kenneth Shepard, James Hone, Solomon Assefa and Dirk Englund, Nat. Photonics, 7, 883(2013).

\bibitem{Gosciniak13} Jacek Gosciniak and Dawn T. H. Tan, Sci. Rep., 3, 1(2013).

\bibitem{RanHao13} Ran Hao, Wei Du, Hongsheng Chen, Xiaofeng Jin, Longzhi Yang, and Erping Li, Appl. Phys. Lett., 103, 061116 (2013).

\bibitem{WeiLi14} Wei Li, Bigeng Chen, Chao Meng, Wei Fang, Yao Xiao, Xiyuan Li, Zhifang Hu, Yingxin Xu, Limin Tong, Hongqing Wang, Weitao Liu, Jiming Bao, and Y. Ron Shen, Nano Lett. 14, 955(2014).

\bibitem{hsu07} Han Hsu and L. E. Reichl, Phys. Rev. B 76, 045418(2007).

\bibitem{hwang07} E. H. Hwang and S. Das Sarma, Phys. Rev. B 75, 205418(2007).

\bibitem{Falkovsky07} L. A. Falkovsky and S. S. Pershoguba, Phys. Rev. B 76, 153410 (2007).

\bibitem{Stauber08} T. Stauber, N. M. R. Peres, and A. K. Geim, Phys. Rev. B 78, 085432 (2008).

\bibitem{Wenhu08} Wenhu Liao, Guanghui Zhou, and Fu Xi, J. Appl. Phys. 104, 126105(2008).

\bibitem{Berahman15} M. Berahman, M. Asad, M. Sanaee, M. H. Sheikhi, Opt. Quant. Electron, 47, 3289(2015).

\bibitem{jablan13} M. Jablan, M. Soljacic, and H. Buljan, Proceedings of the IEEE, 101, 1689(2013).

\bibitem{jablan09} M. Jablan, H. Buljan, and M. Soljacic, Phys. Rev. B., 80, 245435(2009).

\bibitem{Breusing11} M. Breusing, S. Kuehn, T. Winzer, E. Malic, F. Milde, N. Severin, J. P. Rabe, C. Ropers, A. Knorr, and T. Elsaesser, Phys. Rev. B 83, 153410 (2011).

\bibitem{malic11} Ermin Malic, Torben Winzer, Evgeny Bobkin, and Andreas Knorr, Phys. Rev. B 84, 205406(2011).

\bibitem{Sun13} B. Y. Sun and M. W. Wu, New J. Phys., 15, 1(2013).

\bibitem{Torben14} Torben Winzer, Ermin Malic, and Andreas Knorr, Phys. Rev. B 87, 165413 (2013).

\bibitem{Mkrtchian12} H. K. Avetissian, A. K. Avetissian, G. F. Mkrtchian, and Kh. V. Sedrakian, Phys. Rev. B 85, 115443 (2012).

\bibitem{Mkrtchian13} H. K. Avetissian, G. F. Mkrtchian, K. G. Batrakov, S. A. Maksimenko, and A. Hoffmann, Phys. Rev. B 88, 245411 (2013).

\bibitem{sipe11} J. Rioux, Guido Burkard, and J. E. Sipe, Phys. Rev. B 83, 195406 (2011).

\bibitem{sipe15} J. L. Cheng, N. Vermeulen, and J. E. Sipe, Phys. Rev. B 92, 235307 (2015).

\bibitem{kelardeh15} Hamed Koochaki Kelardeh, Vadym Apalkov, and Mark I. Stockman, Phys. Rev. B 91, 045439(2015).



\bibitem{Entin10} M. V. Entin, L. I. Magarill, and D. L. Shepelyansky, Phys. Rev. B, 81, 165441(2010).

\bibitem{Karch10} J. Karch, P. Olbrich, M. Schmalzbauer, C. Zoth, C. Brinsteiner, M. Fehrenbacher, U. Wurstbauer, M. M. Glazov, S. A. Tarasenko, E. L. Ivchenko, D.Weiss, J. Eroms, R. Yakimova, S. Lara-Avila, S. Kubatkin, and S. D. Ganichev, Phys. Rev. Lett., 105, 227402(2010).

\bibitem{Chongyun11} Chongyun Jiang, V. A. Shalygin, V. Yu. Panevin, S. N. Danilov, M. M. Glazov, R. Yakimova, S. Lara-Avila, S. Kubatkin, and S. D. Ganichev, Phys. Rev. B, 84, 125429(2011).

\bibitem{Glazov14} M.M. Glazov and S.D. Ganichev, Physics Reports, 535, 101–138(2014).

\bibitem{kumar16} Anshuman Kumar, Andrei Nemilentsau, Kin Hung Fung, George Hanson, Nicholas X. Fang, and Tony Low, Phys. Rev. B 93, 041413(R)(2016).

\bibitem{rana11} Farhan Rana, Jared H. Strait, Haining Wang, and Christina Manolatou, Phys. Rev. B 84, 045437 (2011).

\bibitem{Nagao01} T. Nagao, T. Hildebrandt, M. Henzler, and S. Hasegawa, Phys. Rev. Lett., 86, 5747(2001).

\bibitem{yuan11} Z. Yuan, Y. Jiang, Y. Gao, M. Kaell, and S. Gao, Phys. Rev. B 83, 165452(2011).


\bibitem{Mikhailov11} S. A. Mikhailov, Phys. Rev. B, 84, 045432(2011).

\bibitem{Smirnova15} D. A. Smirnova, A. E. Miroshnichenko, Y. S. Kivshar, and A. B. Khanikaev, Phys. Rev. B, 92, 161406(R)(2015).

\bibitem{christensen12} Johan Christensen, Alejandro Manjavacas, Sukosin Thongrattanasiri, Frank H. L. Koppens, and F. Javier Garcia de Abajo, ACS Nano, 6, 431(2012).

\bibitem{nikitin11} A. Yu. Nikitin, F. Guinea, F. J. Garcia-Vidal, and L. Martin-Moreno, Phys. Rev. B 84, 161407(R) (2011).

\bibitem{nikitin12} A. Yu. Nikitin, F. Guinea, F. J. Garcia-Vidal, and L. Martin-Moreno, Phys. Rev. B 85, 081405(R) (2012).

\bibitem{koppens11} Frank H. L. Koppens, Darrick E. Chang, and F. Javier Garcia de Abajo, Nano Lett. 11, 3370(2011).

\bibitem{farhat13} Mohamed Farhat, Sebastien Guenneau, and Hakan Bagci, Phys. Rev. Lett. 111, 237404(2013).

\bibitem{schiefele13} Jurgen Schiefele, Jorge Pedros, Fernando Sols, Fernando Calle, and Francisco Guinea, Phys. Rev. Lett. 111, 237405(2013).

\bibitem{Xiaolong13} Xiaolong Zhu, Wei Yan, Peter Uhd Jepsen, Ole Hansen, N. Asger Mortensen, and Sanshui Xiao, Appl. Phys. Lett., 102, 131101(2013).


\bibitem{Sarma13} S. Das Sarma and Qiuzi Li, Phys. Rev. B, 87, 235418(2013).

\bibitem{ashish03} Ashish Kumar Gupta, Ofir E. Alon, and Nimrod Moiseyev, Phys. Rev. B 68, 205101(2003).

\bibitem{qi11} Xiao-Liang Qi and Shou-Cheng Zhang, Rev. Mod. Phys. 83, 1057(2011).








\end{thebibliography}
\end{document}